\documentclass[prc,letterpaper,twocolumn,showpacs,showkeys,lengthcheck,tightenlines,
               nofootinbib,preprintnumbers,superscriptaddress]{revtex4-1}

\usepackage[utf8]{inputenc}

\usepackage{titlesec}
\usepackage{dcolumn}
\usepackage{bm}
\usepackage{amsmath,amssymb,amsfonts}
\allowdisplaybreaks[4]

\usepackage{graphicx}
\usepackage{subfigure}
\usepackage{color}
\usepackage{bbold}

\usepackage{slashed}
\usepackage[svgnames]{xcolor}
\usepackage[english]{babel}
\usepackage{blindtext}
\usepackage{microtype}
\usepackage{tikz}

\usepackage[colorlinks=true,urlcolor=blue,citecolor=blue]{hyperref}
\usepackage{ifpdf}
\usepackage{float}

\graphicspath{{.}{figures/}}

\titlespacing{\section}{5pt}{12pt plus 4pt minus 2pt}{8pt plus 2pt minus 2pt}
\titlespacing{\subsection}{0pt}{12pt plus 4pt minus 2pt}{8pt plus 2pt minus 2pt}

\begin{document}

\title{The kaon off-shell generalized parton distributions and transverse momentum dependent parton distributions}

\author{Jin-Li Zhang}
\email[]{jlzhang@njit.edu.cn}
\affiliation{School of Mathematics and Physics, Nanjing Institute of Technology, Nanjing 211167, China }




\begin{abstract}	
We investigate the off-shell generalized parton distributions (GPDs) {and transverse momentum dependent parton distributions (TMDs)} of kaons within the framework of the Nambu--Jona-Lasinio model, employing proper time regularization. {Compared to the pion case, the off-shell effects in kaons are of similar magnitude, modifying the GPDs by about $10\%$--$25\%$, which is notable. The absence of crossing symmetry leads to odd powers in the $x$-moments of the off-shell GPDs, giving rise to new off-shell form factors.} We analyze the relations among these kaon off-shell form factors by analogy with electromagnetic form factors. Our results extend the off-shell GPDs from pions to kaons and simultaneously address the associated off-shell form factors. We also compare the off-shell and on-shell gravitational form factors of the kaon. In addition, the off-shell kaon TMD shows a stronger dependence on the momentum fraction $x$ than its on-shell counterpart.	
\end{abstract}

\maketitle
\section{Introduction}
One of the central topics in hadronic physics is the understanding of the hadron structure in terms of quark and gluon degrees of freedom. The partonic structure of a hadron is commonly described by light-cone parton distribution functions (PDFs)~\cite{Xu:2024nzp,Wu:2025rto,Tanisha:2025qda}, which correspond to diagonal matrix elements of certain operators and provide crucial inputs for theoretical predictions in deep inelastic scattering (DIS) and other high-energy processes. However, PDFs only give information on the longitudinal momentum distribution of partons. Accessing the full three-dimensional structure requires the study of non-diagonal (off-forward) matrix elements, which are parameterized by GPDs~\cite{Mueller:1998fv,Ji:1996nm,Radyushkin:1997ki,Ji:1998pc,Theussl:2002xp,Diehl:2003ny,Zhang:2020ecj,Zhang:2021mtn,Zhang:2021shm,Zhang:2021tnr,Zhang:2021uak,Zhang:2022zim,Mezrag:2022pqk,Qiu:2022bpq,Zhang:2024adr,Goharipour:2025zsw}.


Since their introduction, GPDs have attracted extensive research interest due to their ability to illuminate partonic probability densities in longitudinal momentum, transverse position, and angular momentum. GPDs thus encode information on how partons are distributed in the plane transverse to the hadron's motion, offering a pathway to reconstruct the three-dimensional structure of hadrons.

On the one hand, different Mellin moments of GPDs are related to various hadronic form factors (FFs)~\cite{Zhang:2024nxl,Bondarenko:2025qch,Hernandez-Pinto:2024kwg,Puhan:2025pfs}, such as electromagnetic FFs~\cite{Cheng:2024cxk}, axial FFs, gravitational FFs (GFFs), and transition FFs~\cite{Zhang:2024dhs}. In particular, GFFs are connected to the energy–momentum tensor~\cite{Ji:2025qax}, enabling a gauge-invariant decomposition of hadron spin. They also provide access to mass and pressure distributions, whose Fourier transforms relate to Breit-frame pressure and shear profiles.

On the other hand, in the forward limit, GPDs reduce to standard PDFs. At zero skewness, a Fourier transform with respect to the transverse momentum transfer yields the impact parameter distribution (IPD), which describes the probability density of finding a parton at the transverse position \(\bm{b}_{\perp}\) relative to the hadron’s center of momentum for a given longitudinal momentum fraction $x$.

Overall, GPDs encapsulate essential information on angular momentum, mass, and mechanical properties of hadrons, offering key insights into the spatial distribution as well as the spin and orbital motion of quarks inside these particles.

These properties make GPDs essential for describing hard exclusive and elastic scattering processes, such as deeply virtual Compton scattering (DVCS)~\cite{Goeke:2001tz,Radyushkin:1996nd,Hobart:2023knc,Xie:2023xkz}, deep virtual meson production (DVMP)~\cite{Muller:2013jur,Favart:2015umi,Cuic:2023mki}, and timelike Compton scattering (TCS)~\cite{Berger:2001xd,Boer:2015fwa,Xie:2022vvl,CLAS:2021lky,Peccini:2021rbt}. These reactions provide key experimental avenues to access GPDs. Recent extractions of GPDs have also been achieved through global analyses of electron scattering data~\cite{Hashamipour:2021kes,Hashamipour:2020kip}.


While the phenomenological extraction of GPDs has not yet reached the maturity of PDFs or FFs, the field is progressing quickly. Key drivers include the advent of next-generation facilities like the U.S. Electron--Ion Collider (EIC) and China’s EicC, together with advances in first-principles computations of GPDs via lattice QCD.


Studies in Refs.~\cite{Aguilar:2019teb,Chavez:2021koz} suggest that pion GPDs could be accessed via the off-shell Sullivan process~\cite{Sullivan:1971kd} at future EIC. A key challenge is that the initial pion is off-shell in this process, making it essential to properly address off-shell effects in the theoretical description of the reaction and the GPDs.


Complementary to GPDs, TMDs~\cite{Zhang:2024plq,Liu:2025fuf} also offer a three-dimensional picture of partonic structure, especially regarding transverse momentum. We thus extend our calculation to include the off-shell TMDs for the kaon.

In Refs.~\cite{Shastry:2023fnc,Broniowski:2023his,Broniowski:2022iip}, the off-shell behavior of pion GPDs was studied in a chiral quark model. In this work, we extend the analysis to kaon off-shell GPDs using the Nambu–Jona-Lasinio (NJL) model~\cite{Klevansky:1992qe,Buballa:2003qv,Zhang:2018ouu,Zhang:2016zto,Cui:2016zqp}. The NJL model preserves the global symmetries of QCD, especially chiral symmetry while integrating  gluonic degrees of freedom to yield point-like quark interactions. This feature, however, renders the model non-renormalizable, making the choice of a regularization scheme essential for its full definition. The NJL approach has been widely applied in studies of hadron structure~\cite{Bentz:1999gx,Noguera:2015iia,Carrillo-Serrano:2015uca,Ceccopieri:2018nop,Freese:2019bhb,Shastry:2022obb,Broniowski:2007si,Bissey:2003yr,RuizArriola:2001rr,Davidson:2001cc,Noguera:2005cc,Volkov:2023pmy,Yu:2023sep} and serves as the basis for our present investigation.


This paper is organized as follows. In Section~\ref{nice}, we begin by briefly introducing the NJL model and then define and calculate the off-shell GPDs of the kaon, also examining their basic properties. The off-shell kaon TMDs are investigated in Section~\ref{tmdoffshell}. Finally, a summary and outlook are provided in Section~\ref{excellent}.

\section{The off-shell GPDs}\label{nice}

\subsection{Nambu–Jona-Lasinio Model}\label{good}

The SU(3) flavor NJL Lagrangian in the $\bar{q}q$ interaction channel is presented in the form described in~\cite{Ishii:1993np},
\begin{align}\label{1}
\mathcal{L}&=\bar{\psi }\left(i\gamma ^{\mu }\partial _{\mu }-\hat{m}\right)\psi+ G_{\pi }[\left(\bar{\psi }\lambda_a\psi\right)^2-\left( \bar{\psi }\gamma _5 \lambda_a \psi \right)^2],
\end{align}
the quark field is represented by the flavor components $\psi^T = (u, d, s)$. {The matrices $\lambda_a$, where $a=1,...,8$, denote the eight Gell--Mann matrices in color space, and $\lambda_0$ = $\sqrt{2/3}$ $ \mathbf{1}_f$, where $\mathbf{1}_f$ is the unit matrix in the flavor space.} The current quark mass matrix is represented as $\hat{m}=\text{diag}\left(m_u,m_d,m_s\right)$. The parameter $G_{\pi}$ denotes the effective coupling strength associated with the scalar interaction channel ($\bar{q}\lambda_a q$) and the pseudoscalar interaction channel ($\bar{q}\gamma_5 \lambda_a q$)~\cite{Cloet:2014rja,Hutauruk:2018zfk}.

\begin{figure}
	\centering
	\includegraphics[clip,width=0.94\linewidth]{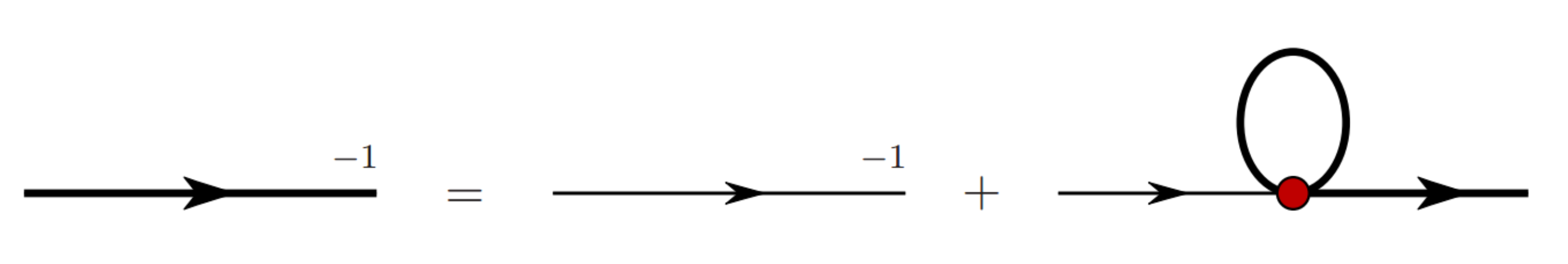}
	\caption{\label{GAP} The NJL gap equation in the Hartree--Fock approximation is illustrated in the figure, where a thin line denotes the elementary quark propagator and a shaded circle represents the  $\bar{q}q$ interaction kernel. This kernel does not incorporate higher-order contributions, such as those from meson loops.}
\end{figure}

The dressed quark propagator within the framework of the NJL model is derived by solving the gap equation illustrated in Fig. \ref{GAP}
\begin{align}\label{2}
S_q(k)=\frac{1}{{\not\!k}-M_q+i \varepsilon},
\end{align}
the dressed quark mass is denoted as \( M_q = (M_u, M_d, M_s) \). The local interaction kernel of the gap equation in Figure \ref{GAP} leads to a constant dressed quark mass \( M_q \), satisfying: 
\begin{align}\label{ncogap}
M_q=m_q+12 i G_{\pi}\int \frac{d^4l}{(2 \pi )^4}\text{Tr}_D[S_q(l)],
\end{align}
the trace is taken over Dirac indices. {It is notable that, unlike the SU(2) case, flavor mixing is absent in SU(3) flavor~\cite{Klevansky:1992qe,Carrillo-Serrano:2016igi}. } Additionally, dynamical chiral symmetry breaking occurs only for $G_{\pi} > G_{critical}$, resulting in a nonzero dressed quark mass $M_q > 0$.

Mesons in the NJL model are treated as $\bar{q}q$ bound states, obtained as solutions to the Bethe--Salpeter equation (BSE). The BSE solution in a given meson channel is described by a channel-dependent two-body $t$-matrix. For example, the reduced $t$-matrices for kaons take the following form: 
\begin{align}\label{3}
\tau_K(q)=\frac{-2i\,G_{\pi}}{1+2G_{\pi}\Pi_{PP}(q^2)},
\end{align}
where the bubble diagram $\Pi_{PP}(q^2)$ is defined as~\cite{Ninomiya:2014kja,Hutauruk:2016sug}
\begin{align}\label{ppbb}
\Pi_{PP}(q^2)\delta_{ij}=3i\int \frac{d^4k}{(2\pi)^4}\text{Tr}[\gamma^5\tau_iS_u(k)\gamma^5\tau_jS_s(k+q)],
\end{align}
where trace is taken over Dirac indices. The mass of the kaon is determined by the pole in the reduced $t$-matrix
\begin{align}\label{5}
1+2\,G_{\pi}\Pi_{PP}(q^2=m_K^2)=0.
\end{align}
Expanding the complete $t$-matrix around the pole yields the homogeneous Bethe--Salpeter vertex for the kaon
\begin{align}\label{6C}
\Gamma_K^{i}=\sqrt{Z_K}\gamma_5\lambda_i,
\end{align}
and the normalization factor is defined as follows:
\begin{align}\label{qmcc}
Z_K^{-1}&=-\frac{\partial}{\partial q^2}\Pi_{PP}(q^2)|_{q^2=m_K^2}.
\end{align}
This residue corresponds to the square of the effective meson--quark--quark coupling constant. The homogeneous Bethe--Salpeter vertex functions play a key role in processes such as triangle diagrams, which are used to determine meson form factors.

The NJL model is non-renormalizable and thus requires regularization. Here we adopt the proper time regularization (PTR) scheme~\cite{Ebert:1996vx,Hellstern:1997nv,Bentz:2001vc}.
\begin{align}\label{4}
\frac{1}{X^n}&=\frac{1}{(n-1)!}\int_0^{\infty}d\tau\, \tau^{n-1}e^{-\tau X}\nonumber\\
& \rightarrow \frac{1}{(n-1)!} \int_{1/\Lambda_{\text{UV}}^2}^{1/\Lambda_{\text{IR}}^2}d\tau \,\tau^{n-1}e^{-\tau X},
\end{align}
where the product of propagators connected via Feynman parametrization is Wick-rotated to Euclidean space, with the resulting denominator denoted as $X$. PTR is then applied to evaluate the final result. $\Lambda_{\text{UV}}$ denotes the ultraviolet cutoff. To account for the absence of confinement in the NJL model, an infrared cutoff of order $\Lambda_{\text{QCD}}$ is employed; we accordingly choose $\Lambda_{\text{IR}} = 0.240$ GeV.

For the dressed quark masses, \( M_u = M_d = 0.4 \) GeV and \( M_s =0.611 \) GeV. The ultraviolet cutoff \( \Lambda_{\text{UV}} \) and the coupling constant \( G_{\pi} \) are fixed by matching empirical values of the pion decay constant and pion mass. The kaon is described as a relativistic bound state of a dressed quark and a dressed antiquark, whose properties are obtained by solving the Bethe--Salpeter equation in the pseudoscalar \( \bar{q}q \) channel. {The full set of parameters used in this work, taken from Ref.~\cite{Hutauruk:2018zfk}, is listed in Table \ref{tb1}.}

An explicit \( Q^2 \) scale is absent in our approach—a typical limitation in most model-based determinations of quark distributions. This necessitates setting a model scale \( Q_0^2 \) by comparing with empirical data. Here we adopt \( Q_0^2 = 0.16 \) GeV\(^2\), which is consistent with Ref.~\cite{Cloet:2007em} and typical for models dominated by valence contributions~\cite{Schreiber:1991tc,Mineo:1999eq,Mineo:2002bg,Ninomiya:2017ggn}.

The following sections make use of the $\mathcal{C}$ functions and corresponding formulas provided in the Appendix. 

\begin{center}
\begin{table}
\caption{The parameter set utilized in our study is presented here. The current quark mass and regularization parameters are expressed in units of GeV, while the coupling constants are measured in units of GeV$^{-2}$. The decay constants are in units of GeV.}\label{tb1}
\begin{tabular}{p{1.0cm} p{1.0cm} p{1.0cm}p{1.0cm}p{1.0cm} p{1.0cm}p{1.0cm}}
\hline\hline
$\Lambda_{\text{IR}}$&$\Lambda _{\text{UV}}$&$m_u$&$m_s$&$G_{\pi}$&$m_K$&$Z_K$\\
\hline
0.240&0.645&0.017&0.321&19.0&0.47&20.47\\
\hline\hline
\end{tabular}
\end{table}
\end{center}

\subsection{The Off-shell Kaon GPDs }\label{qq}

\begin{figure}[t]
	\centering
	\includegraphics[clip,width=0.94\linewidth]{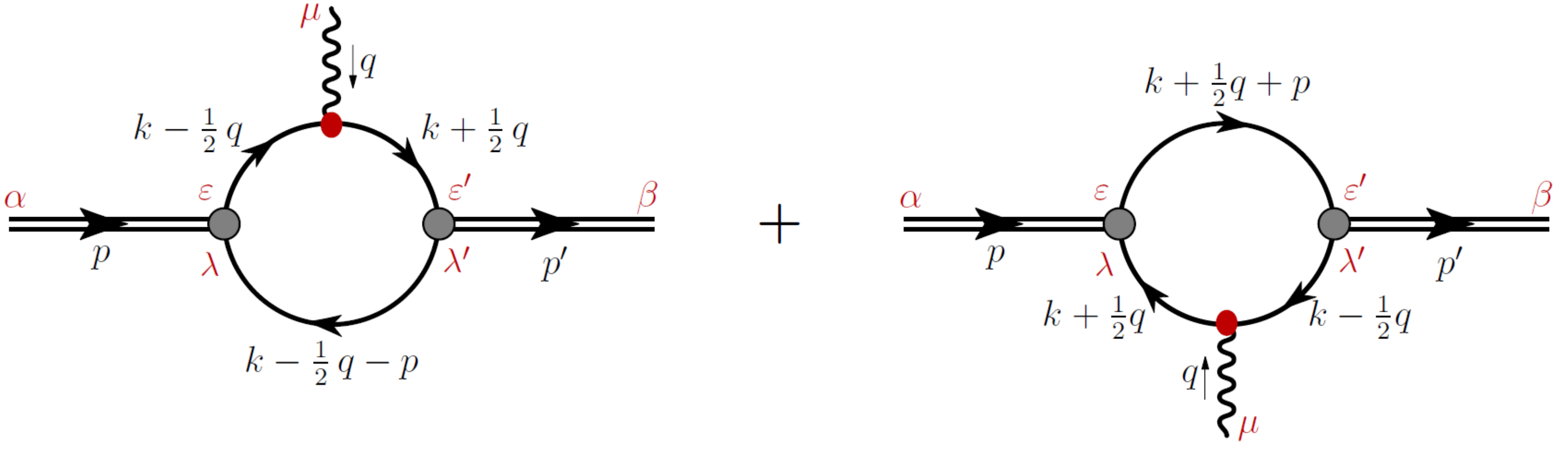}
	\caption{\label{GPD} Diagrams of off-shell GPDs for kaons, where \( p^2 \neq p^{\prime 2} \neq m_K^2 \).}
\end{figure}

GPDs are currently undergoing extensive examination across a variety of models and theoretical frameworks~\cite{Mezrag:2023nkp,Siddikov:2025tjo}. Within the NJL model, the kaon off-shell GPDs are depicted in Figure \ref{GPD}, where $p$ and $p^{\prime}$ represent the initial and final kaon momenta, respectively. For the off-shell case with \( p^2 \neq p^{\prime 2} \neq m_K^2 \), the relevant kinematic variables are defined as follows: 
\begin{align}\label{4}
t=q^2=(p^{\prime}-p)^2=-Q^2,
\end{align}
\begin{align}\label{5}
P=\frac{p+p^{\prime}}{2}, \quad \xi=\frac{p^+-p^{\prime+}}{p^++p^{\prime+}},\quad n^2=0,
\end{align}
where $\xi$ represents the skewness parameter, and $n$ denotes the light-cone four-vector defined as $n=(1,0,0,-1)$ in the context of light-cone coordinates
\begin{align}\label{4A}
v^{\pm}&=(v^0\pm v^3), \quad  \mathbf{v}=(v^1,v^2),
\end{align}
for any four-vector \( v^+ \), it can be defined in light-cone coordinates as follows:
\begin{align}\label{4B}
v^+=v\cdot n.
\end{align}
The vector and tensor quark GPDs of kaon are defined as~\cite{Shastry:2023fnc} 
\begin{align}\label{dgpd}
H^q(x,\xi,t,p^2,p^{\prime 2})&=\frac{1}{2}\int \frac{dz^-}{2\pi}e^{\frac{i}{2}x(p^++p^{\prime+})z^-}\nonumber\\
&\times \langle p^{\prime }|\bar{q}(-\frac{1}{2}z)\gamma^+q(\frac{1}{2}z)|p\rangle \mid_{z^+=0,\bm{z}=0},
\end{align}
\begin{align}\label{dtgpd}
&\quad \frac{P^+q^j-P^jq^+}{P^+m_K}E^q(x,\xi,t,p^2,p^{\prime 2})\nonumber\\
&=\frac{1}{2}\int \frac{dz^-}{2\pi}e^{\frac{i}{2}x(p^++p^{\prime+})z^-}\nonumber\\
&\times \langle p^{\prime }|\bar{q}(-\frac{1}{2}z)i\sigma^{+j}q(\frac{1}{2}z)|p\rangle \mid_{z^+=0,\bm{z}=0},
\end{align}
where \( x \) represents the longitudinal momentum fraction. The function \( H^q(x,\xi,t,p^2,p^{\prime2}) \) denotes the $u$ quark non-spin-flip or vector GPD, while \( E^q(x,\xi,t,p^2,p^{\prime2}) \) corresponds to the $u$ quark spin-flip or tensor GPD.

The operators depicted in Figure \ref{GPD} for off-shell kaon GPDs are structured as follows:
\begin{align}\label{6A}
	\textcolor{red}{\bullet}_1 &=\gamma^+\delta(x-\frac{k^+}{P^+}),
\end{align}
\begin{align}\label{6A}
	\textcolor{red}{\bullet}_2 &=i\sigma^{+j} \delta(x-\frac{k^+}{P^+}),
\end{align}
where the first term corresponds to the vector GPD and the second to the tensor GPD.

In the NJL model, the vector and tensor GPDs for the $u$ quark in the $K^+$ meson are defined as follows: 
\begin{align}\label{gpddd}
H^u\left(x,\xi,t,p^2,p^{\prime2}\right)
&=2i N_c Z_K\int \frac{d^4k}{(2 \pi )^4}\delta_n^x (k)\nonumber\\
&\times \text{Tr}\left[\gamma _5 S_u \left(k_{+}\right)\gamma ^+ S_u\left(k_{-}\right)\gamma _5 S_s\left(k_P\right)\right],
\end{align}
\begin{align}\label{tgpddd}
&\quad \frac{P^+q^j-P^jq^+}{P^+m_K} E^u\left(x,\xi,t,p^2,p^{\prime2}\right)\nonumber\\
&=2i N_c Z_K\int \frac{d^4k}{(2 \pi )^4}\delta_n^x (k) \nonumber\\
&\times \text{Tr}\left[\gamma _5 S _u\left(k_{+}\right)i\sigma^{+j} S_u\left(k_{-}\right)\gamma _5 S_s\left(k_P\right)\right],
\end{align}
where $\delta_n^x (k)=\delta (xP^+-k^+)$ $k_+=k+\frac{q}{2}$, $k_-=k-\frac{q}{2}$, $k_P=k-P$.

We adopt the following notation:
%
\begin{align}\label{a6}
	\mathcal{D}_{k_{+}}^u&=\left(k+\frac{q}{2}\right)^2-M_u^2,
\end{align}
\begin{align}\label{a6}
	\mathcal{D}_{k_{-}}^u&=\left(k-\frac{q}{2}\right)^2-M_u^2,
\end{align}
\begin{align}\label{a6}
	\mathcal{D}_{k_P}^s&=\left(k-p-\frac{q}{2}\right)^2-M_s^2,
\end{align}
%
one can derive the following simplified formulas
\begin{align}
	p\cdot q&=\frac{p^{\prime2}-p^2-q^2}{2},
\end{align}
\begin{align}
	k\cdot q&=\frac{1}{2} \left(\mathcal{D}_{k_{+}}^u-\mathcal{D}_{k_{-}}^u\right),
\end{align}
\begin{align}
	k\cdot p&=-\frac{1}{2} \left(\mathcal{D}_{k_P}^s-\mathcal{D}_{k_{-}}^u-M_u^2+M_s^2-\frac{p^{\prime2}+p^2-q^2}{2}\right),
\end{align}
\begin{align}
	k^2&=\frac{1}{2} \left(\mathcal{D}_{k_{+}}^u+\mathcal{D}_{k_{-}}^u\right)+M_u^2-\frac{q^2}{4},
\end{align}
after some calculation we arrive at
\begin{align}\label{agpdf1}
&\quad H^u\left(x,\xi,t,p^2,p^{\prime2}\right)\nonumber\\
&=\frac{ N_c Z_K }{8 \pi ^2}\left[\theta_{\bar{\xi} 1}\bar{\mathcal{C}}_1(\sigma_4)+\theta_{\xi 1}\bar{\mathcal{C}}_1(\sigma_5)+\frac{\theta_{\bar{\xi} \xi} }{\xi }x \bar{\mathcal{C}}_1(\sigma_6)\right]\nonumber\\
&+\frac{N_c Z_K }{8\pi ^2}\int_0^1 d\alpha\frac{\theta_{\alpha \xi}}{\xi}\frac{1}{\sigma_7}\bar{\mathcal{C}}_2(\sigma_7)\nonumber\\
&\times ((p^{\prime 2}-p^2)\xi+(1-x)t+x (p^2+p^{\prime 2})-2x\left(M_u-M_s\right)^2),
\end{align}
\begin{align}\label{agtpdf}
&E^u\left(x,\xi,t,p^2,p^{\prime 2}\right)\nonumber\\
=&\frac{N_c Z_K }{4\pi ^2 }\int_0^1 d\alpha \frac{\theta_{\alpha \xi}m_K}{\xi}((M_s-M_u)\alpha+M_u)\frac{\bar{\mathcal{C}}_2(\sigma_7)}{\sigma_7},
\end{align}
and
\begin{align}
	\theta_{\bar{\xi} 1}&=x\in[-\xi, 1],
\end{align}
\begin{align}
	\theta_{\xi 1}&=x\in[\xi, 1],
\end{align}
\begin{align}
	\theta_{\bar{\xi} \xi}&=x\in[-\xi, \xi],
\end{align}
\begin{align}
	\theta_{\alpha \xi}&=x\in[\alpha (\xi +1)-\xi , \alpha  (1-\xi)+\xi ]\cap x\in[-1,1].
\end{align}
We denote the step function by $\theta$. It takes the value of $1$ in the corresponding region and is otherwise equal to zero. These results pertain to the region where $\xi > 0$. Under the transformation $\xi \rightarrow -\xi$, we observe that $\theta_{\bar{\xi} 1} \leftrightarrow \theta_{\xi 1}$; furthermore, both $\theta_{\bar{\xi} \xi}/\xi$ and $\theta_{\alpha \xi}/\xi$ remain invariant.

Here, we present only the off-shell kaon GPDs for the $u$ quark. For the $s$ quark, we can derive the GPDs using the relationship outlined below.
\begin{align}\label{rs}
	H^u\left(x,\xi,t,p^2,p^{\prime 2}\right)&=-H^s\left(-x,\xi,t,p^2,p^{\prime 2},M_u\leftrightarrow M_s\right),
\end{align}		
\begin{align}\label{rs}		
	E^u\left(x,\xi,t,p^2,p^{\prime 2}\right)&=-E^s\left(-x,\xi,t,p^2,p^{\prime 2},M_u\leftrightarrow M_s\right),
\end{align}
where \( M_u \leftrightarrow M_s \) indicates the exchange of \( M_u \) with \( M_s \).

\begin{figure}[t]
\includegraphics[clip,width=0.94\linewidth]{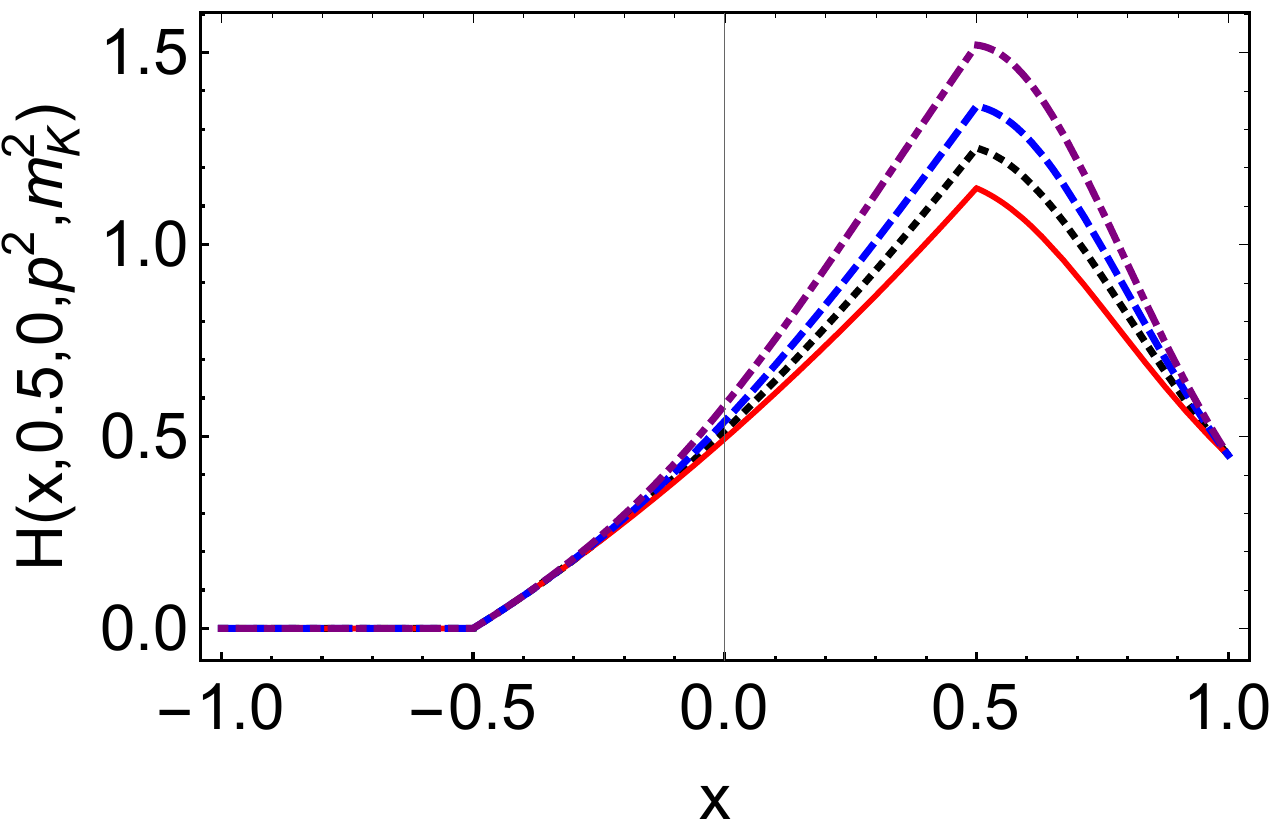}
\includegraphics[clip,width=0.94\linewidth]{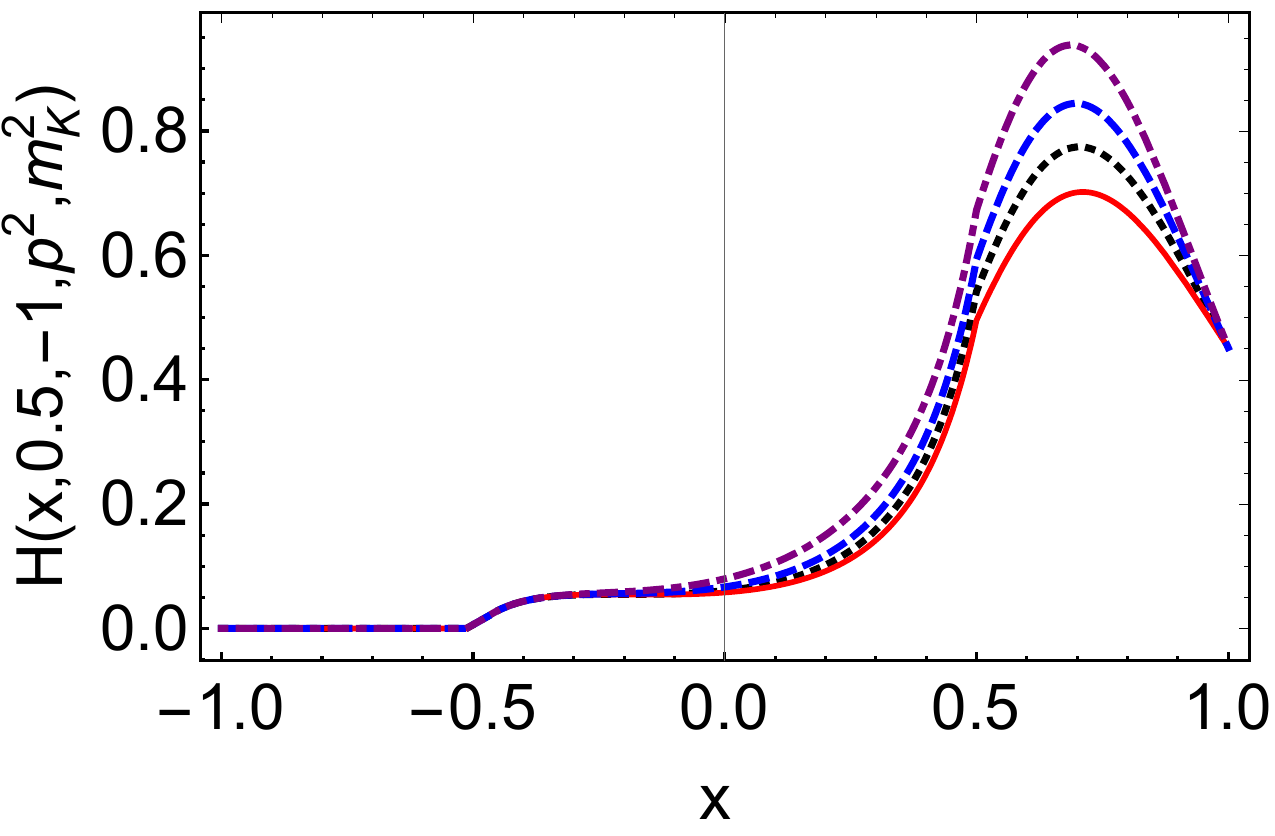}
\caption{\label{gpddgpd}
Kaon off-shell vector GPD: \( H^u(x,\xi,t,p^2,m_K^2) \) in Equation (\ref{agpdf1}) with $\xi=0.5$ for
$t = 0$ (\textbf{upper panel}) and $t = -1$ (\textbf{lower panel}). Case $p^2 = 0$ is represented by the red solid line, \(p^2 = m_K^2\) black dotted line, $p^2 = 0.4$ GeV$^2$ blue dashed line, and $p^2 = 0.6$ GeV$^2$ purple dot-dashed line.
}
\end{figure}

\begin{figure}[t]
\includegraphics[clip,width=0.94\linewidth]{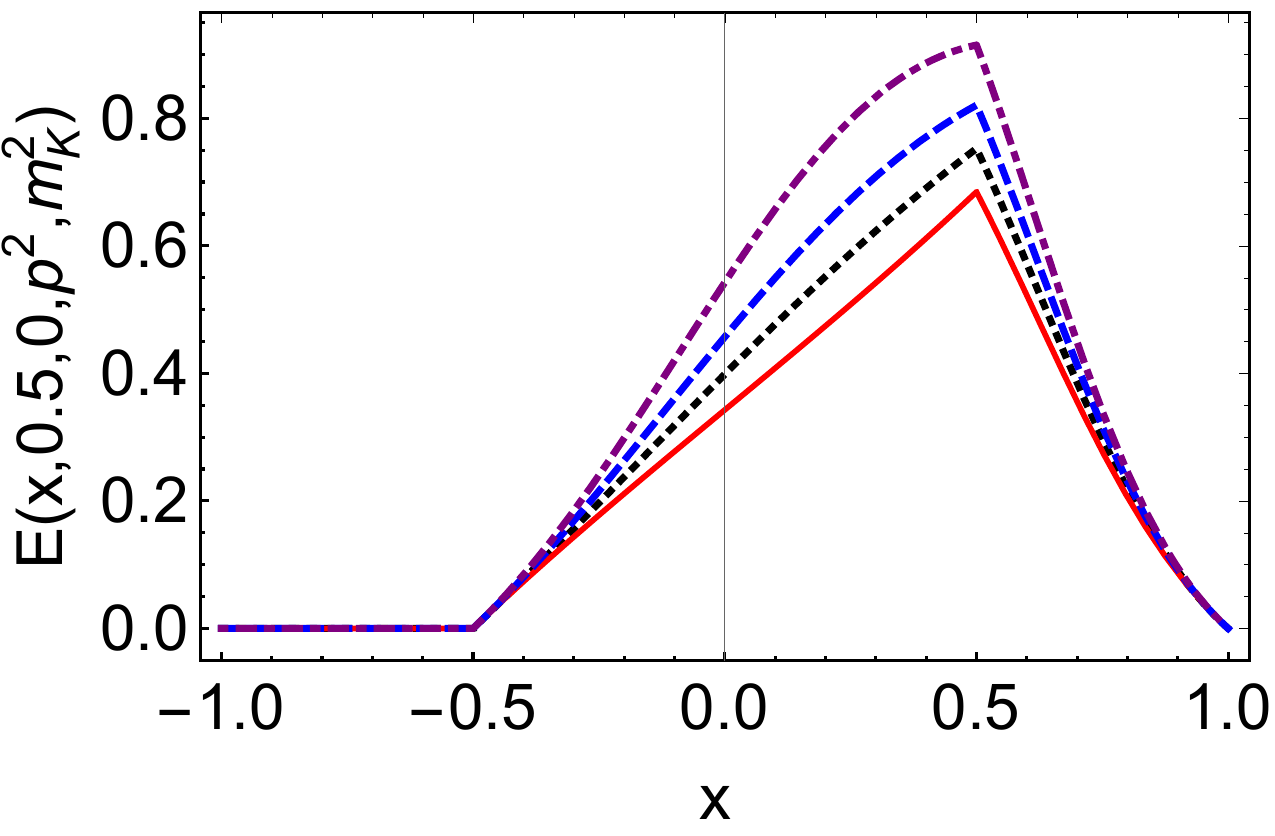}
\includegraphics[clip,width=0.94\linewidth]{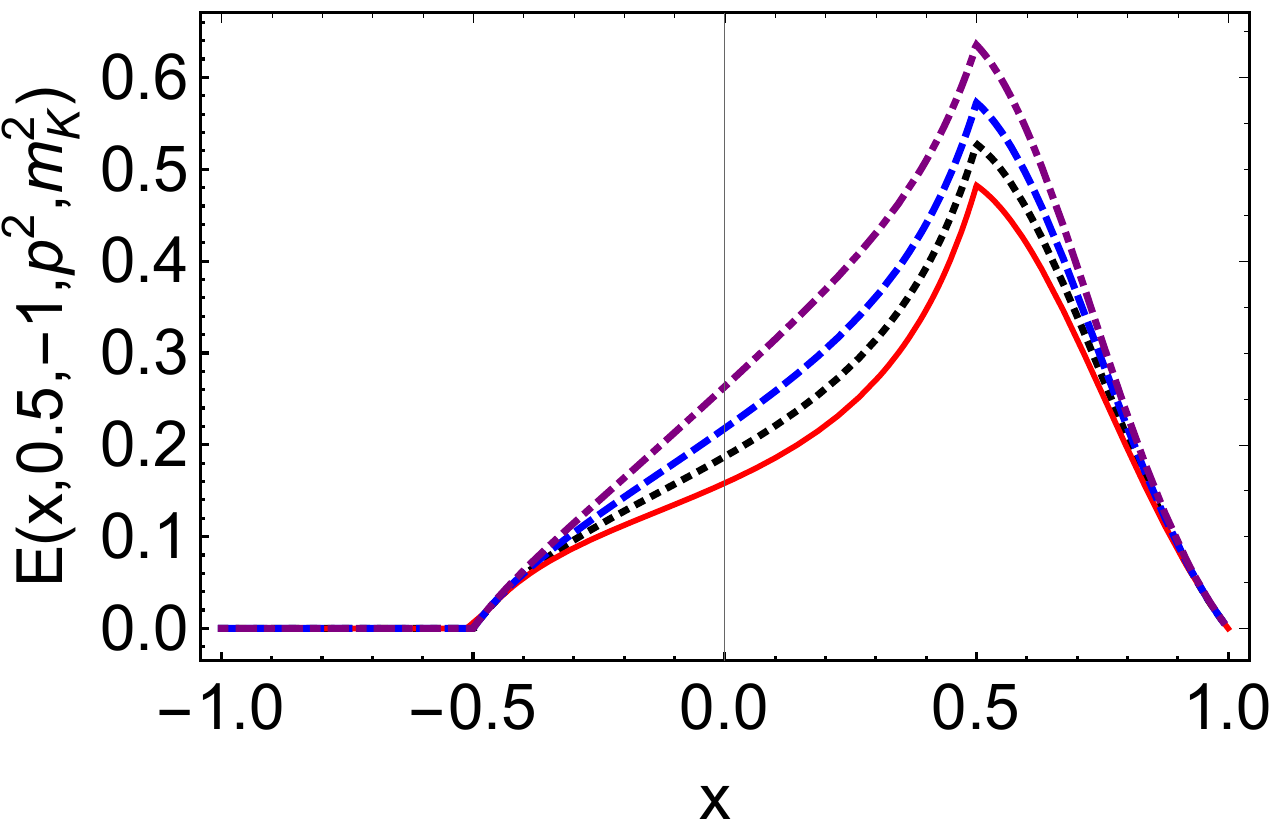}
\caption{\label{tgpddgpd}
Kaon off-shell tensor GPD: \( E^u(x,\xi,t,p^2,m_K^2) \) in Equation (\ref{agtpdf}) with $\xi=0.5$ for $t = 0$ (\textbf{upper panel}) and $t = -1$ (\textbf{lower panel}). Case $p^2 = 0$ is represented by the red solid line, \(p^2 = m_K^2\) black dotted line, $p^2 = 0.4$ GeV$^2$ blue dashed line, and $p^2 = 0.6$ GeV$^2$ purple dot-dashed line.
}
\end{figure}

The distributions of \( H^u(x,\xi,t,p^2,m_K^2) \) and \( E^u(x,\xi,t,p^2,m_K^2) \) are shown in Figures \ref{gpddgpd} and \ref{tgpddgpd}, where the off-shell dependence arises from varying \( p^2 \) while \( p^{\prime 2}=m_K^{2} \) is held fixed. The results demonstrate that as \( p^2 \) increases, the off-shell effects in the half-off-shell kaon GPDs grow significantly. In particular, at \( p^{2}=0.4 \) GeV$^2$, the relative deviation reaches about $10\%$, and increases to roughly $25\%$ at \( p^{2}=0.6 \) GeV$^2$. As expected, the magnitude of the off-shell effects depends on \( p^2 \), and increases gradually with \( p^2 \). The strongest effects are observed near \( x = \xi \). These results are consistent with earlier findings for off-shell pion GPDs reported in Refs.~\cite{Broniowski:2022iip,Shastry:2023fnc,Zhang:2025xtn}, suggesting that off-shell effects in GPDs can be substantial in physical processes and should be properly accounted for.


In Ref.~\cite{Zhang:2025xtn}, the off-shell pion GPDs were studied using the NJL model. Given that the kaon mass is larger than the pion mass, different \( p^2 \) values were chosen for comparison: for the pion, $p^2=0.0,0.2,0.4$ GeV\(^2\), and for the kaon, $p^2=0.0,0.4,0.6$ GeV\(^2\), where $m_K^2=0.495^2=0.245$ GeV$^2$. If \( p^2 \) is increased by the same amount relative to each meson’s on-shell mass—e.g., by $0.2$ GeV$^2$ —the off-shell effects become comparable: for the pion at \( p^2 =(0.14^2+ 0.2) \) GeV\(^2\) and for the kaon at \( p^2 = (0.495^2+0.2) \) GeV\(^2\). This suggests that the off-shell effects in pions and kaons are of similar magnitude, a conclusion consistent with the results from the chiral quark model in Ref.~\cite{Broniowski:2022iip}.

\subsection{The Properties Of Kaon off-shell GPDs}\label{qq}

\subsubsection{Forward Limit}\label{qqQ}
In the case where the initial and final kaon momenta are equal, $p=p^{\prime}$, implying $\xi=0$ and $t=0$, the vector GPD reduces to the kaon PDF,
\begin{align}\label{hpdf}
& \quad u_K(x,p^2)\nonumber\\
&=\frac{3Z_K}{4\pi ^2}\bar{\mathcal{C}}_1(\sigma_1)\nonumber\\
&+\frac{3Z_K}{2\pi ^2} x(1-x) (p^2-\left(M_u-M_s\right)^2) \frac{\bar{\mathcal{C}}_2(\sigma_1)}{\sigma_1},
\end{align}
when the skewness parameter $\xi=0$, the function $H^u(x,0,0,p^2,p^2)$ vanishes for $x\in [-1,0]$. In contrast, for $x\in [0,1]$, the off-shell PDF of the $u$-quark in the kaon differs from that in the pion, as described in the NJL model~\cite{Zhang:2021shm,Zhang:2025xtn}.

Thus far, we have analyzed off-shell effects in GPDs. We now turn to the off-shell behavior of the normalization factor given in Equation~(\ref{qmcc})~\cite{Shastry:2023fnc}. When \( p^2 \neq m_K^2 \), the value of \( Z_K \) changes accordingly. Several representative values of \( Z_K \) for different choices of  \( p^2 \) are listed in Table~\ref{tb2}. 

The kaon PDF should satisfy the relationship
\begin{align}\label{nm}
	\int_0^1u_K\left(x,p^2\right) dx =1,
\end{align}
where distinct values of $p^2$ correspond to different values of $Z_K$.

\begin{figure}
	\centering
	\includegraphics[width=0.47\textwidth]{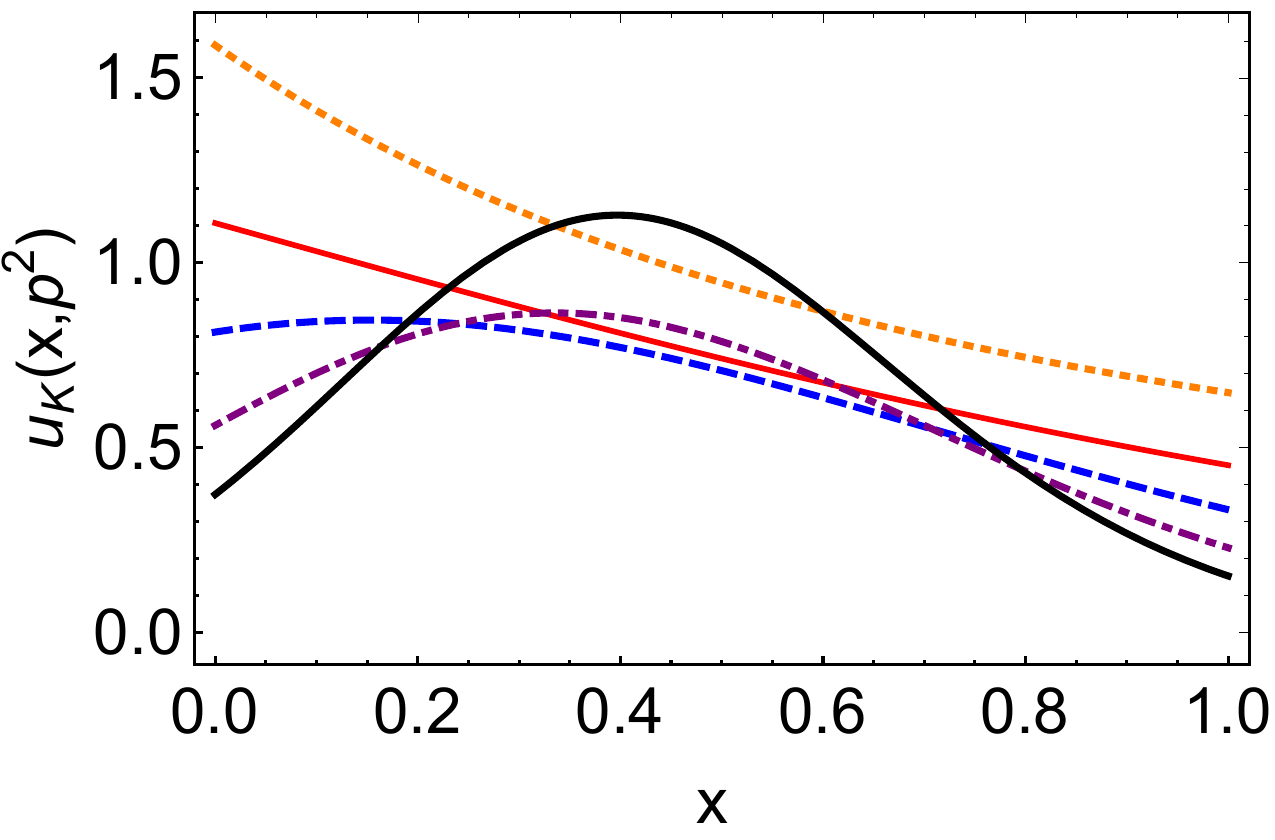}
	\caption{Off-shell	$u$ quark PDFs in the kaon: $u_K\left(x,p^2\right)$, for different virtualities: $p^2=0$ GeV$^{2}$---orange dotted curve, $p^2=m_K^2=0.495^2$ GeV$^{2}$---red solid curve, $p^2=0.4$ GeV$^{2}$---blue dashed curve, $p^2=0.6$ GeV$^{2}$---purple dot-dashed curve, and $p^2=0.8$ GeV$^{2}$---black solid thick curve.}\label{offshellpdf}
\end{figure}

Figure \ref{offshellpdf} shows the off-shell \( u \)-quark PDF in the kaon. When \( p^2 = 0 \) or \( p^2 = m_K^2 \), the distribution is nearly flat in $x$. As \( p^2 \)
increases—especially in the off-shell case—a stronger $x$-dependence develops. While the NJL model captures low-energy QCD features well, at large \( Q^2 \) the PDF must be evolved, which will further modify its $x$-dependence. Unlike the pion PDF, the peak of the kaon PDF shifts to $x<0.5$, breaking the symmetry around $x = 0.5$ and distinguishing it from the pion case. Additionally, for the $u$-quark PDF in the kaon, the value at $x=0$ exceeds that at $x=1$. As $p^2$ increases, both endpoint values decrease.

\begin{center}
	\begin{table}
		\caption{The normalization factor \(Z_K\) corresponding to various values of \(p^2\), where \(p^2\) is expressed in units of GeV$^2$.}\label{tb2}
		\begin{tabular}{p{1.0cm} p{1.2cm} p{1.2cm}p{1.2cm}p{1.2cm} p{1.2cm}}
			\hline\hline
			$p^2$&$0$&$0.495^2$&$0.4$&$0.6$&$0.8$\\
			\hline
			\(Z_K\)&31.006&20.906&16.048&11.148&7.497\\
			\hline\hline
		\end{tabular}
	\end{table}
\end{center}

\subsubsection{Polynomiality Condition}\label{Bqq}
The $x$-moments of the off-shell GPDs also include odd powers of the skewness parameter $\xi$.
%
\begin{align}\label{aF41}
	\int_{-1}^1 x^n H^q(x,\xi,t,p^2,p^{\prime 2}) dx  &= \sum _{i=0}^{(n+1)} A_{n,i}^u(t,p,p^{\prime})\xi^i,
\end{align}
\begin{align}\label{aF42}
	\int_{-1}^1 x^n E^q(x,\xi,t,p^2,p^{\prime 2})dx &= \sum _{i=0}^{(n+1)} B_{n,i}^q(t,p^2,p^{\prime 2})\xi^i.
\end{align}
%
This highlights the remarkable polynomiality property of GPDs: the sums corresponding to the $x$-integrals of $x^nH^q$ and $x^nE^q$ are polynomials in $\xi$ of order $n+1$.

When $n=0$, we obtain FFs  
\begin{subequations}
\begin{align}\label{ptc}
\int_{-1}^1 x^0 H^u dx  &= A_{1,0}^u+\xi A_{1,1}^u=F^u-G^u\xi\,, \\
\int_{-1}^1 x^0 E^u dx  &= B_{1,0}^u+\xi B_{1,1}^u,
\end{align}
\end{subequations}
where $ A_{1,0}^u $ and $B_{1,0}^u$ are the vector and tensor FFs of the $u$ quark, while $A_{1,1}^u$ and $B_{1,1}^u$ are additional FFs induced by off-shell effects. The former are symmetric under $p \leftrightarrow p^{\prime}$, whereas the latter are antisymmetric.

The functions $F(t,p^2,p^{\prime2})$ and $G(t,p^2,p^{\prime2})$ are defined in the general covariant structure of the kaon-photon vertex:
\begin{align}
	\Gamma^{\mu}(p,p^{\prime})=2P^{\mu}F(t,p^2,p^{\prime2})+q^{\mu}G(t,p^2,p^{\prime2}),
\end{align}
at $t=0$, the relationship~\cite{Broniowski:2022iip}
\begin{align}\label{g}
	G(t,p^2,p^{\prime2})=\frac{(p^{\prime2}-p^2)}{t}\left[F(0,p^2,p^{\prime2})-F(t,p^2,p^{\prime2})\right],
\end{align}
where $G(0,p^2,p^{\prime2})=-(p^{\prime2}-p^2)dF(t,p^2,p^{\prime2})/dt|_{t=0}$ implies, via crossing symmetry (time-reversal invariance)~\cite{Broniowski:2023his}, that \( G(t,p^2,p^2) = 0 \). Crossing symmetry further requires that on-shell GPDs are invariant under interchange of initial and final momenta  ($p \leftrightarrow p^{\prime}$), leading to \( H(x,\xi,t) = H(x,-\xi,t) \). Here, \( F^u(t,p^2,p^{\prime2}) \) denotes the $u$-quark electromagnetic FF, with the normalization \( F^u(0,m_K^2,m_K^2) = 1 \). From the off-shell kaon GPDs, one can derive   
\begin{align}\label{aF3}
&\quad A_{1,0}^u (Q^2,p^2,p^{\prime 2})\nonumber\\
&=\frac{N_cZ_K}{8\pi ^2}\int_0^1 dx \, \bar{\mathcal{C}}_1(\sigma_1)+\frac{N_cZ_K}{8\pi ^2}\int_0^1 dx \, \bar{\mathcal{C}}_1(\sigma_2)\nonumber\\
&+\frac{N_cZ_K}{4\pi ^2}\int_0^1dx\int_0^{1-x}dy\frac{1}{\sigma_8}\bar{\mathcal{C}}_2(\sigma_8)\nonumber\\
&\times ((1-x-y)((p^2+p^{\prime 2})-2(M_u-M_s)^2)-(x+y)Q^2),
\end{align}
\begin{align}\label{aF3}
&\quad A_{1,1}^u (Q^2,p^2,p^{\prime2})\nonumber\\
&=\frac{N_cZ_K}{8\pi ^2}\int_0^1 dx \, \bar{\mathcal{C}}_1(\sigma_1)-\frac{N_cZ_K}{8\pi ^2}\int_0^1 dx \, \bar{\mathcal{C}}_1(\sigma_2)\nonumber\\
&-\frac{N_cZ_K}{4\pi ^2}\int_0^1dx\int_0^{1-x}dy(p^2-p^{\prime2})\frac{\bar{\mathcal{C}}_2(\sigma_8)}{\sigma_8} \nonumber\\
&-\frac{N_cZ_K}{4\pi ^2}\int_0^1dx\int_0^{1-x}dy\frac{\bar{\mathcal{C}}_2(\sigma_8)}{\sigma_8} \nonumber\\
&\times ((p^2-p^{\prime 2})+(y-x)(p^2+p^{\prime 2}+Q-2(M_u-M_s)^2)),
\end{align}
\begin{align}\label{aF3}
&\quad B_{1,0}^u (Q^2,p^2,p^{\prime2})\nonumber\\
&=\frac{N_cZ_K }{2\pi^2}\int _0^1dx \int_0^{1-x} dy \nonumber\\
&\times m_K \left((M_s-M_u)(1-x-y)+M_u\right) \frac{\bar{\mathcal{C}}_2(\sigma_8)}{\sigma_8},
\end{align}
where $B_{1,1}^u(Q^2,p^2,p^{\prime2})=0$.

\begin{figure}
	\centering
	\includegraphics[width=0.47\textwidth]{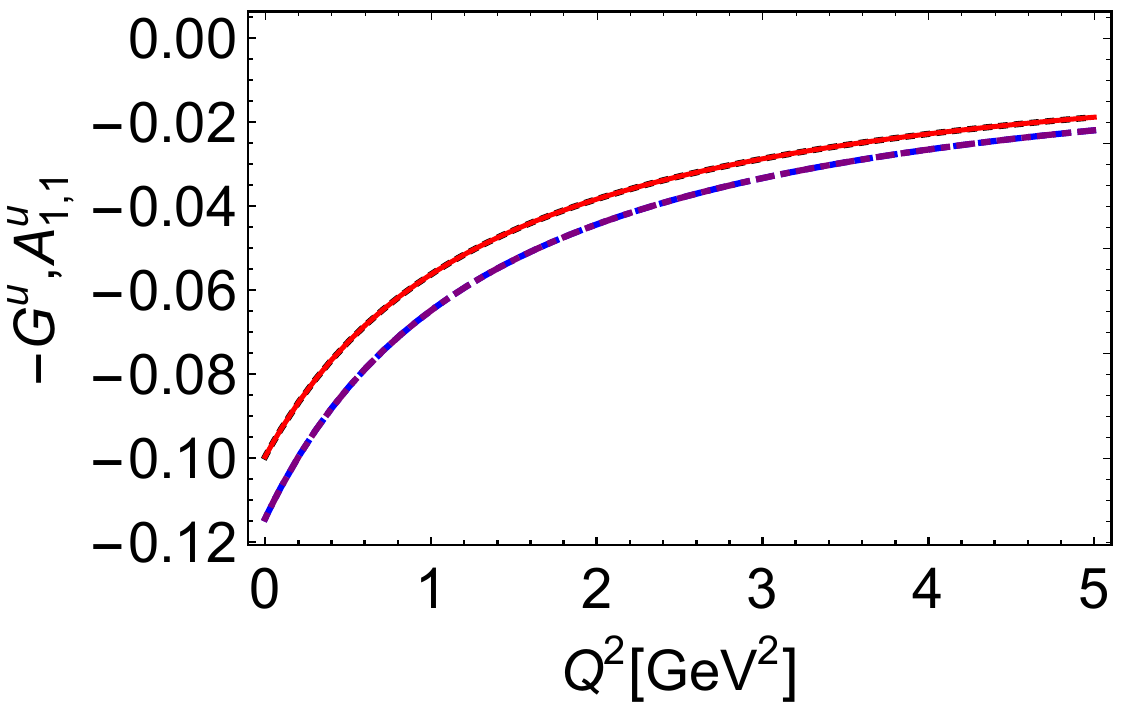}
	\caption{The graphs of \( A_{1,1}^u (Q^2,p^2,p^{\prime 2}) \) and \( -G^u (Q^2,p^2,p^{\prime 2}) \): \( -G^u (Q^2,0.2,0.3) \)---black dotted curve,  \( A_{1,1}^u (Q^2,0.2,0.3) \)---red solid curve, \( -G^u (Q^2,0.3,0.4) \)---blue dashed curve, and \( A_{1,1}^u (Q^2,0.3,0.4) \)---purple dot-dashed curve.}\label{ag}
\end{figure}

We show that our result for \( A_{1,1}^u (Q^2,p^2,p^{\prime 2}) \) equals \( -G^u (Q^2,p^2,p^{\prime 2}) \) as given in Equation~(\ref{g}), in agreement with Ref.~\cite{Broniowski:2022iip}. Figure~\ref{ag} displays \( A_{1,1}^u (Q^2,p^2,p^{\prime 2}) \) above and \( -G^u (Q^2,p^2,p^{\prime 2}) \) in Equation (\ref{g}), confirming that the two curves coincide exactly.

For $n=1$, GPDs should follow the sum rule:
\begin{subequations}
\begin{align}\label{ab94}
&\quad \int_{-1}^1 x\, dx H^u(x,\xi,t,p^2,p^{\prime2})\nonumber\\
&= A_{2,0}^u(t,p^2,p^{\prime2})+\xi A_{2,1}^u (t,p^2,p^{\prime2})+\xi^2 A_{2,2}^u (t,p^2,p^{\prime2})\nonumber\\
&=\theta_2^u(t,p^2,p^{\prime2})-\xi\theta_3^u(t,p^2,p^{\prime2})-\xi^2\theta_1^u(t,p^2,p^{\prime2})\,, \\
&\quad \int_{-1}^1 x\, dx E^u(x,\xi,t,p^2,p^{\prime 2})\nonumber\\
&= B_{2,0}^u(t,p^2,p^{\prime 2})+\xi B_{2,1}^u(t,p^2,p^{\prime2})+\xi^2 B_{2,2}^u(t,p^2,p^{\prime2}),
\end{align}
\end{subequations}
where $\theta_2^u$ pertains to the mass distribution of the $u$ quark within the kaon, while $\theta_1^u$ is associated with the pressure distribution of the $u$ quark. {The polynomial contains the terms $\xi^0$, $\xi^1$ and $\xi^2$, thereby satisfying Equations (\ref{aF41})-(\ref{aF42}).}  The generalized FFs for $n=1$ are as follows:
\begin{align}\label{ag20}
&A_{2,0}^u(Q^2,p^2,p^{\prime2})\nonumber\\
=&\frac{N_c Z_K}{8 \pi ^2}\int_0^1dx  x \bar{\mathcal{C}}_1(\sigma_1)+\frac{N_c Z_K}{8 \pi ^2}\int_0^1dx  x \bar{\mathcal{C}}_1(\sigma_2)  \nonumber\\
+&\frac{N_c Z_K}{4\pi ^2}  \int_0^1 dx \int_0^{1-x} dy(1-x-y) \frac{1}{\sigma_8}\bar{\mathcal{C}}_2(\sigma_8)\nonumber\\
\times & ( ((p^2+p^{\prime2})-2(M_s-M_u)^2) (1-x-y)-Q^2  (x+y)),
\end{align}
\begin{align}\label{theta31}
&A_{2,1}=\frac{N_c Z_K}{8 \pi ^2} \left(\frac{1}{p^2}- \frac{1}{p^{\prime 2}} \right)(\mathcal{C}_0(M_u^2)-\mathcal{C}_0(M_s^2)) \nonumber\\
&+\frac{N_cZ_K}{8\pi ^2}\int_0^1 dx\bar{\mathcal{C}}_1(\sigma_1)\nonumber\\
&\times \left(2 x-1 +\frac{2 x(p^2+p^{\prime 2}-2(M_u-M_s)^2)}{Q^2 }-\frac{M_s^2-M_u^2}{p^2}\right)\nonumber\\
&-\frac{N_cZ_K}{8\pi ^2}\int_0^1 dx\bar{\mathcal{C}}_1(\sigma_2)\nonumber\\
&\times \left(2 x-1 +\frac{2 x(p^2+p^{\prime 2}-2(M_u-M_s)^2)}{Q^2 }-\frac{M_s^2-M_u^2}{p^{\prime 2}}\right)\nonumber\\
&+\frac{N_c Z_K}{2\pi^2}\int_0^1 dx \int_0^{1-x} dy(1-x-y)(p^{\prime 2}-p^2)\frac{\bar{\mathcal{C}}_2(\sigma_8)}{\sigma_8}\nonumber\\
&\times \left((x+y)-\frac{(p^{\prime2}+p^2-2(M_u-M_s)^2) (1-x-y)}{Q^2}\right) ,
\end{align}
\begin{align}\label{a93}
&A_{2,2}=-\frac{N_cZ_K }{2\pi^2}\int_0^1 dx x (1-2 x)\bar{\mathcal{C}}_1(\sigma_3)\nonumber\\
&+\frac{N_cZ_K }{8\pi^2}\int_0^1 dx(\bar{\mathcal{C}}_1(\sigma_2)-\bar{\mathcal{C}}_1(\sigma_1))(p^2-p^{\prime 2})\nonumber\\
&\times \left(\frac{x-1}{Q^2 }+\frac{x (p^{\prime 2}+p^2-2(M_u-M_s)^2)}{Q^4 }\right)\nonumber\\
&+\frac{N_cZ_{\pi } }{8\pi^2}\int_0^1 dx(\bar{\mathcal{C}}_1(\sigma_2)+\bar{\mathcal{C}}_1(\sigma_1))\nonumber\\
&\times (1-x)\frac{(p^{\prime 2}+p^2-2(M_u-M_s)^2)}{Q^2}\nonumber\\
&-\frac{N_c Z_K}{4\pi^2}\int_0^1 dx \int_0^{1-x} dy\bar{\mathcal{C}}_1(\sigma_8)\nonumber\\
&\times \left(\frac{(p^{\prime 2}+p^2-2(M_u-M_s)^2)}{Q^2 }+1\right)\nonumber\\
&+\frac{N_c Z_K}{4\pi^2}\int_0^1 dx \int_0^{1-x} dy(1-x-y)(p^{\prime 2}-p^2)^2\frac{\bar{\mathcal{C}}_2(\sigma_8)}{\sigma_8}\nonumber\\
&\times \left(\frac{(p^{\prime2}+p^2-2(M_u-M_s)^2)(1-x-y)}{Q^4}-\frac{  (x+y)}{Q^2}\right),
\end{align}
\begin{align}\label{a93}
&B_{2,0}^u(Q^2,p^2,p^{\prime2})\nonumber\\
=&\frac{N_c Z_K  }{2\pi ^2} \int_0^1 dx \int_0^{1-x} dy \frac{1}{\sigma_8}\bar{\mathcal{C}}_2(\sigma_8) \nonumber\\
\times &  m_K ((1-x-y)^2 (M_s-M_u)+(1-x-y)M_u),
\end{align}
\begin{align}\label{aF3}
&\quad B_{2,1}^u (Q^2,p^2,p^{\prime2})\nonumber\\
&=-\frac{N_cZ_K}{4\pi ^2}\int_0^1 dx \frac{m_K(x(M_s-M_u)+M_u)}{ Q^2} \bar{\mathcal{C}}_1(\sigma_1)\nonumber\\
&+\frac{N_cZ_K}{4\pi ^2}\int_0^1 dx \frac{m_K(x(M_s-M_u)+M_u)}{ Q^2}\bar{\mathcal{C}}_1(\sigma_2)\nonumber\\
&-\frac{N_cZ_K }{2\pi^2}\int _0^1dx \int_0^{1-x} dy (1-x-y)\frac{\bar{\mathcal{C}}_2(\sigma_8)}{\sigma_8}\nonumber\\
&\times m_K ((1-x-y) (M_s-M_u)+M_u)\frac{(p^{\prime2}-p^2)}{Q^2} .
\end{align}
For the $u$ quark tensor GPD $E^u(x,\xi,t,p^2,p^{\prime2})$ of the kaon within the NJL model, we find that $B_{2,2}^u(Q^2,p^2,p^{\prime2})=0$.  

\subsubsection{Impact Parameter Dependent PDFs}
The impact parameter dependent PDFs are given by,
\begin{align}\label{aG9}
&\quad q\left(x,\bm{b}_{\perp}^2,p^2,p^{\prime2}\right)\nonumber\\
&=\int \frac{d^2\bm{q}_{\perp}}{(2 \pi )^2}e^{-i\bm{b}_{\perp}\cdot \bm{q}_{\perp}} H^q\left(x,0,-\bm{q}_{\perp}^2,p^2,p^{\prime2}\right).
\end{align}
This implies that the impact parameter dependent PDFs introduced above are given by the Fourier transform of $H^u\left(x,0,-\bm{q}_{\perp}^2\right)$, thus, once $H^u\left(x,0,-\bm{q}_{\perp}^2\right)$ is known, the parton distribution as a function of the transverse position $\bm{b}_{\perp}$ and momentum fraction $x$ can be determined. Similarly, the off-shell tensor impact parameter dependent PDF is obtained as the Fourier transform of \( E^u\left(x,0,-\bm{q}_{\perp}^2\right) \).

When $\xi \rightarrow 0$ and $t\neq 0$, the off-shell GPDs reduce to
\begin{align}\label{aG91}
&\quad H^u\left(x,0,-\bm{q}_{\perp}^2,p^2,p^{\prime2}\right)\nonumber\\
&=\frac{3 Z_K}{8 \pi ^2} (\bar{\mathcal{C}}_1(\sigma_1)+\bar{\mathcal{C}}_1(\sigma_2))\nonumber\\
&+\frac{3 Z_K}{4 \pi ^2}\int_0^{1-x} d\alpha  \frac{1}{\sigma_9}\bar{\mathcal{C}}_2(\sigma_9)\nonumber\\
&\times ((x-1)\bm{q}_{\perp}^2+x ( (p^{\prime2}+p^2)-2(M_u-M_s)^2)),
\end{align}
\begin{align}\label{aG911}
&\quad E^u\left(x,0,-\bm{q}_{\perp}^2,p^2,p^{\prime2}\right)\nonumber\\
&=\frac{N_cZ_K}{2\pi ^2}\int_0^{1-x} d\alpha \,m_K ((M_s-M_u)\alpha+M_u) \frac{\bar{\mathcal{C}}_2(\sigma_9)}{\sigma_9},
\end{align}
where \( x \) is defined within the interval \( x \in [0, 1] \), we can derive the following:
\begin{widetext}
\begin{align}\label{1ipspdf}
u_K\left(x,\bm{b}_{\perp}^2,p^2,p^{\prime2}\right)
&=\frac{N_cZ_K}{8\pi ^2} \int \frac{d^2\bm{q}_{\perp}}{(2 \pi )^2}e^{-i\bm{b}_{\perp}\cdot \bm{q}_{\perp}} (\bar{\mathcal{C}}_1(\sigma_1) +\bar{\mathcal{C}}_1(\sigma_2))\nonumber\\
&+\frac{N_cZ_K}{32\pi ^3}\int_0^{1-x} d\alpha \int d\tau \left(\frac{(x-1)+ \alpha  \tau  x (1-\alpha -x) ((p^{\prime2}+p^2)-2(M_u-M_s)^2)}{\alpha ^2 \tau ^2 (1-\alpha -x)^2}+\frac{(x-1)\bm{b}_{\perp}^2}{4 \alpha ^3 \tau ^3 (1-\alpha -x)^3}\right)\nonumber\\
&\times e^{-\tau  \left((1-x) M_u^2+x M_s^2-(x\alpha p^2+x(1-\alpha-x) p^{\prime 2})\right)}e^{- \frac{\bm{b}_{\perp}^2}{4\tau \alpha (1-\alpha -x)}},
\end{align}
\begin{align}\label{2ipspdf}
u_K^T\left(x,\bm{b}_{\perp}^2,p^2,p^{\prime2}\right)
&=\frac{N_cZ_K}{16\pi^3}\int_0^{1-x}  d\alpha \int d\tau  \frac{m_K((M_s-M_u)\alpha+M_u) }{\alpha\left(1-\alpha-x\right)\tau } e^{- \frac{1}{4\tau (1-\alpha -x)\alpha }\bm{b}_{\perp}^2}e^{-\tau  \left((1-x) M_u^2+x M_s^2-(x\alpha p^2+x(1-\alpha-x) p^{\prime 2})\right)},
\end{align}
\end{widetext}
integrating $u_K\left(x,\bm{b}_{\perp}^2,p^2,p^{\prime2}\right)$ over $\bm{b}_{\perp}$ yields the $u$-quark PDF given in Equation~(\ref{hpdf}). The quantity \( u_K^T\left(x,\bm{b}_{\perp}^2,p^2,p^{\prime 2}\right) \) denotes the off-shell tensor impact parameter dependent PDF derived from the tensor GPD \( E^u \). 
\begin{figure}
	\centering
	\includegraphics[width=0.47\textwidth]{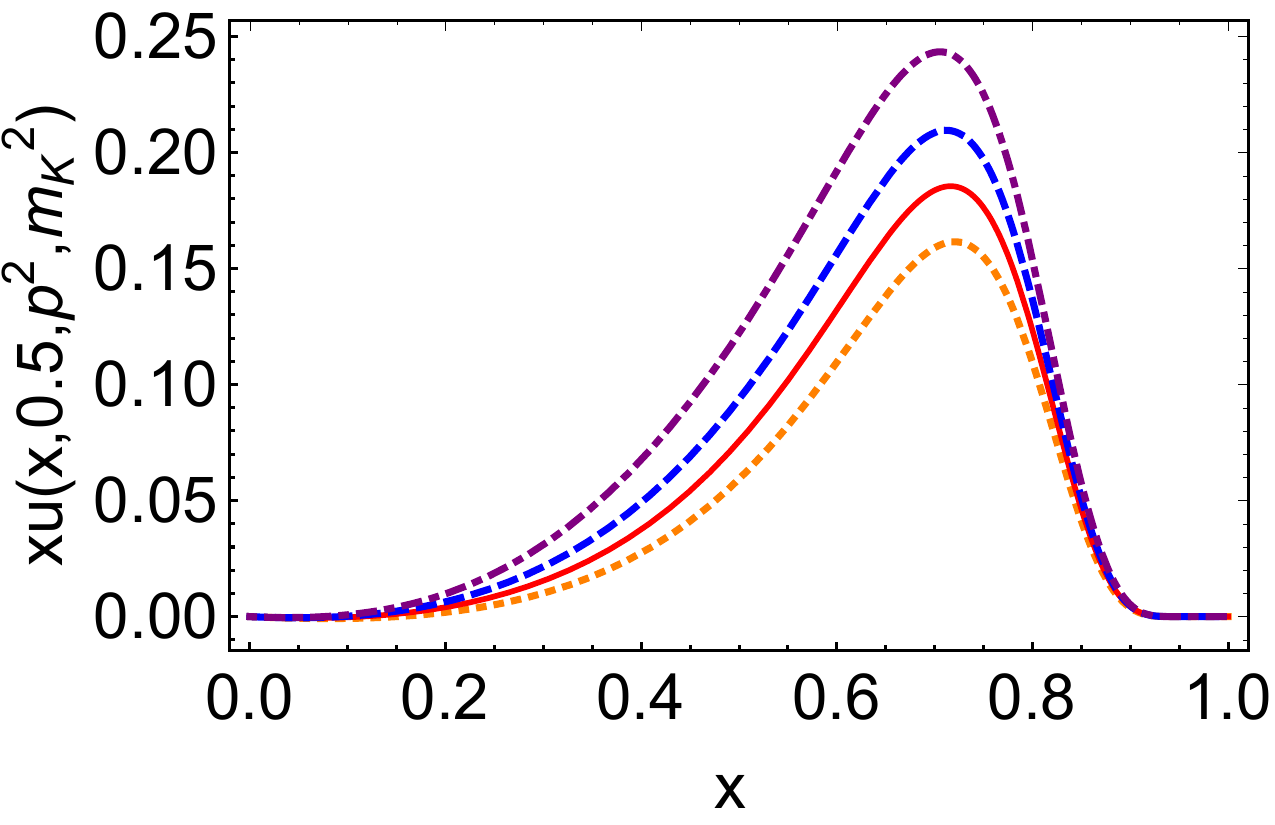}
	\qquad
	\includegraphics[width=0.47\textwidth]{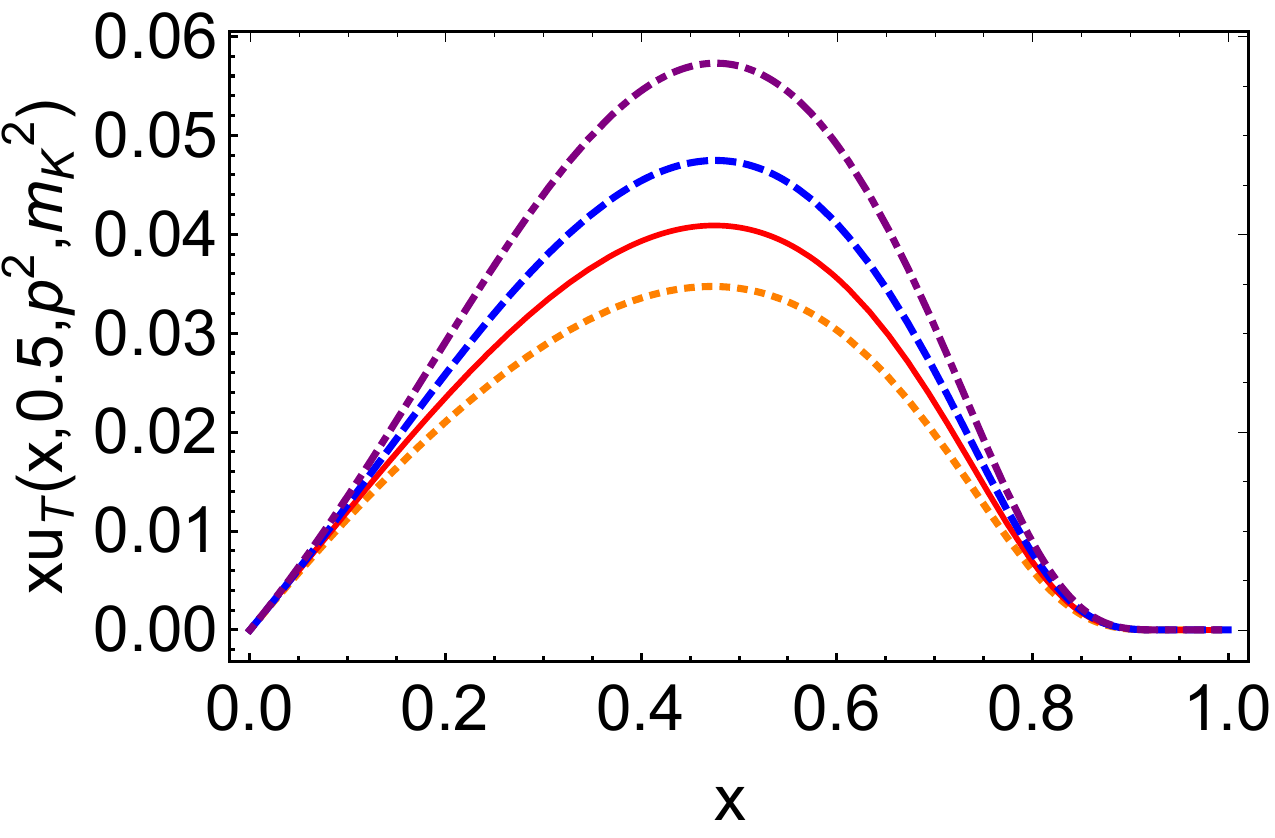}
	\caption{Impact parameter space PDFs: $x u\left(x,0.5,p^2,m_K^2\right)$ (\textbf{upper panel}); the $\delta^2(\bm{b}_{\perp})$ component in the first line of Equation (\ref{1ipspdf}) is suppressed in the image; $x u_T\left(x,0.5,p^2,m_K^2\right)$ (\textbf{lower panel}); both panels with $p^2=0$ GeV$^{2}$---orange dotted curve; $p^2=0.495^2$ GeV$^{2}$---red solid curve; $p^2=0.4$ GeV$^{2}$---blue dashed curve; $p^2=0.6$ GeV$^{2}$---purple dot-dashed curve.}\label{qxb2}
\end{figure}

Figure \ref{qxb2} shows the off-shell impact parameter dependent PDFs multiplied by $x$, evaluated at $\bm{b}_{\perp}^2=0.5$ GeV$^{-2}$ for different values of $p^2$. The maximum of $xu\left(x,0.5,p^2,m_K^2\right)$ occurs at a larger $x$ than that of $xu_T\left(x,0.5,p^2,m_K^2\right)$. As $p^2$ increases, the peak heights of both distributions grow. At the same time, the peak position of $xu$ shifts to smaller $x$, though it remains above \( x = 0.5 \). In contrast, the peak position of $xu_T$ stays near \( x \approx 0.5 \) as $p^2$ varies.

\section{The kaon off-shell TMDs}\label{tmdoffshell}
TMDs have been widely investigated~\cite{Abdulov:2022itv,Loomis:2025vtk}, yet studies specifically addressing their off-shell counterparts are still scarce.

\begin{figure}
	\centering
	\includegraphics[width=0.47\textwidth]{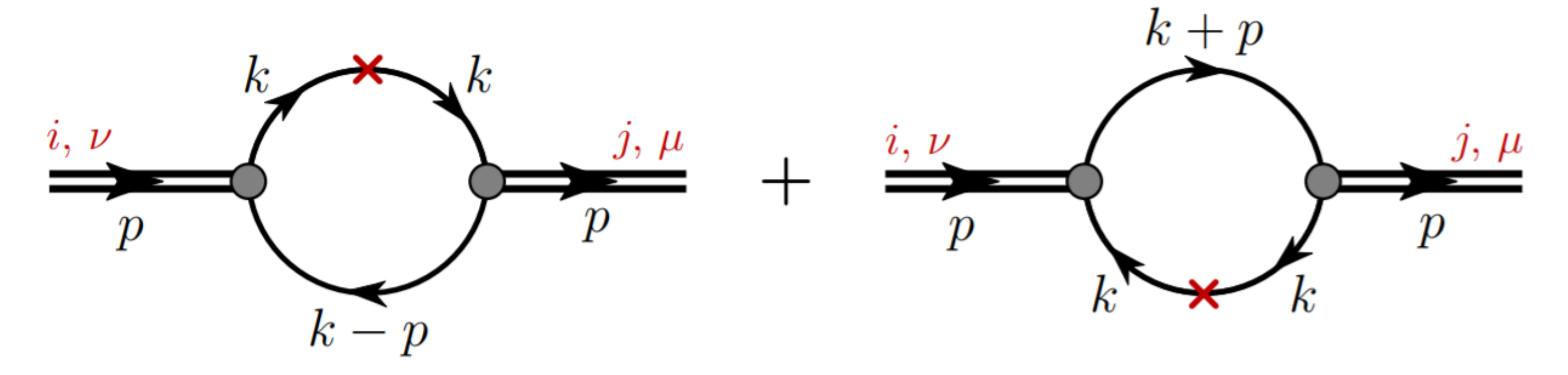}
	\caption{Feynman diagrams illustrating the kaon TMDs within the NJL model are presented. The shaded circles denote the kaon Bethe-Salpeter vertex functions, while the solid lines represent the dressed quark propagator. The operator insertion takes the form $\gamma^+ \delta(x-\frac{k^+}{p^+})$. The left diagram corresponds to the TMDs of the up $u$ quark, whereas the right diagram pertains to those of the strange $s$ quark in relation to the kaon.}\label{TMD}
\end{figure}

\begin{figure}
	\centering
	\includegraphics[width=0.47\textwidth]{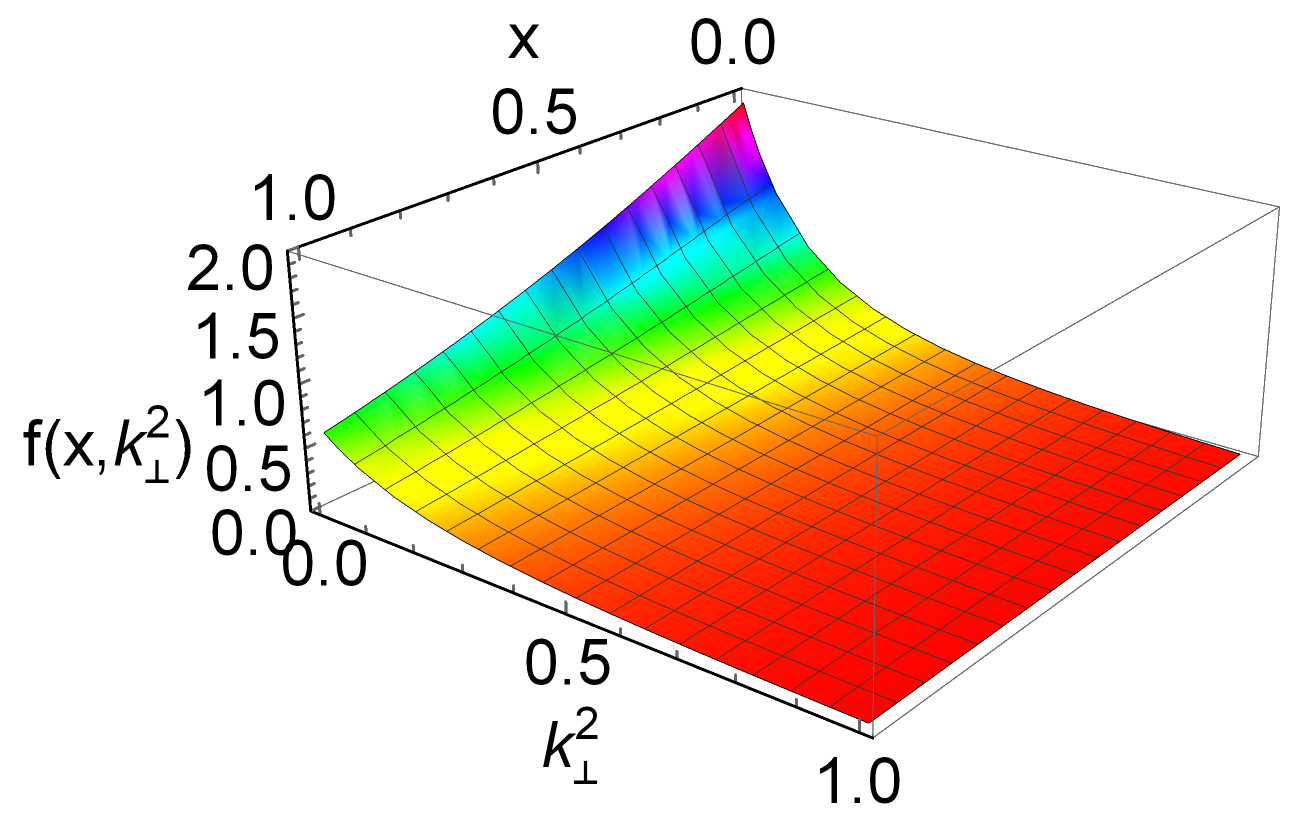}
	\qquad
	\includegraphics[width=0.47\textwidth]{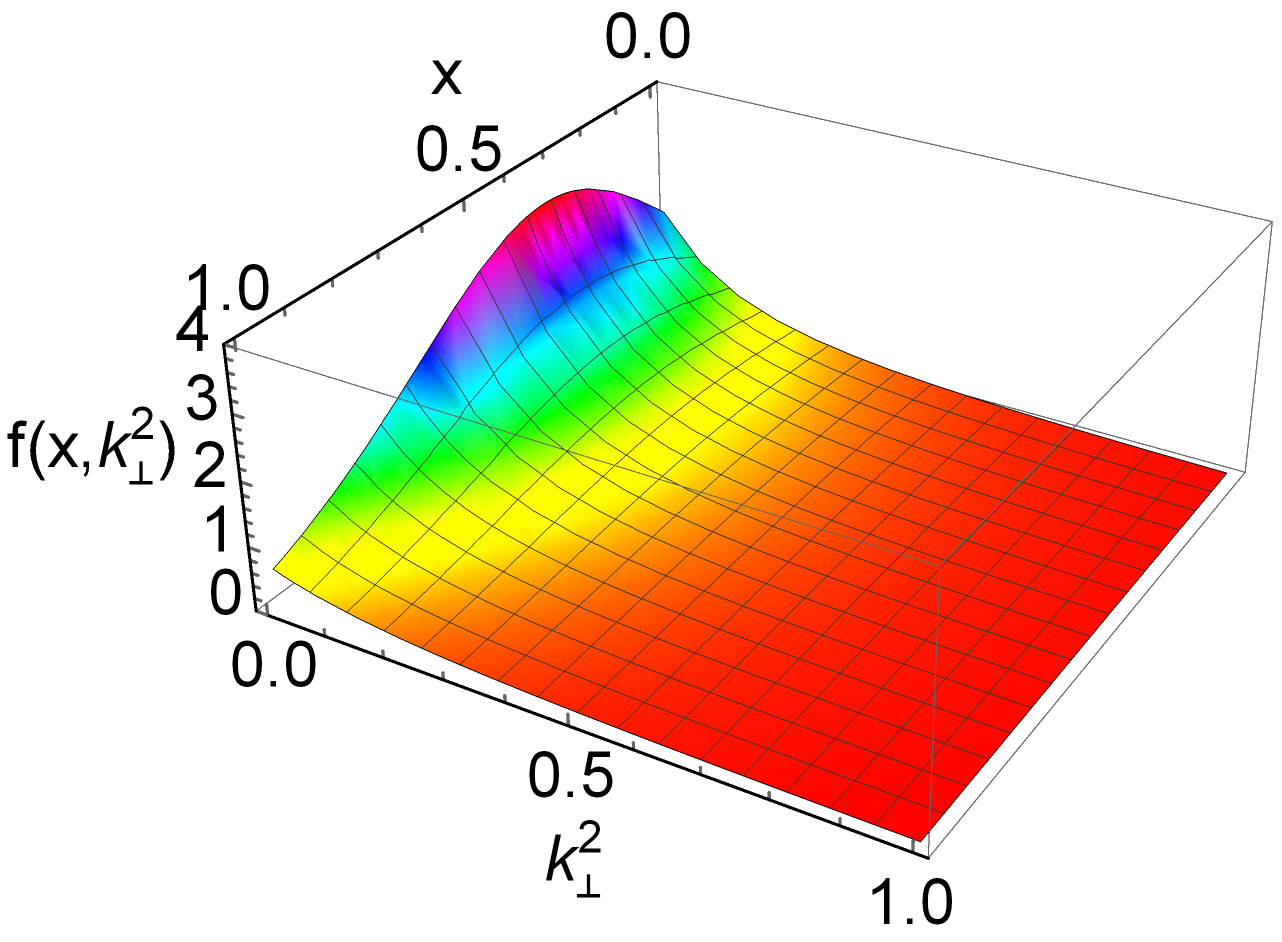}
	\caption{Kaon TMDs: The on-shell kaon TMDs $f^u(x,\bm{k}_{\perp}^2,m_K^2)$ (\textbf{upper panel}); The off-shell kaon TMDs $f^u(x,\bm{k}_{\perp}^2,0.6)$ (\textbf{lower panel}).}\label{tmd1}
\end{figure}

Figure \ref{TMD} shows the kaon TMD within the NJL model. Its definition, analogous to that of the pion TMD~\cite{Noguera:2015iia}, is given by:
\begin{align}\label{gpddd1}
	&\langle\Gamma\rangle^u(x,\bm{k}_{\perp}^2)=-\frac{i N_c Z_K}{p^+}\int \frac{\mathrm{d}k^+\mathrm{d}k^-}{(2 \pi )^4}\delta (x-\frac{k^+}{p^+})\nonumber\\
	&\times  \text{tr}_{\text{D}}\left[\gamma^5 S_u \left(k\right)\gamma^+S_u\left(k\right)\gamma^5 S_s\left(k-p\right)\right],
\end{align}
where $\text{tr}_{\text{D}}$ represents a trace over spinor indices. As a result, we have successfully derived the final expression for the off-shell kaon TMD.
\begin{align}\label{aG91}
	&\quad f^u(x,\bm{k}_{\perp}^2,p^2)\nonumber\\
	&=\frac{N_cZ_K}{2\pi ^3} \frac{\bar{\mathcal{C}}_2(\sigma_{10})}{\sigma_{10}}\nonumber\\
	&+\frac{N_cZ_K}{4\pi ^3}  x(1-x)( p^2-(M_u-M_s)^2)\frac{6\bar{\mathcal{C}}_3(\sigma_{10})}{\sigma_{10}^2},
\end{align}
Figure \ref{tmd1} displays the three-dimensional distributions of the on-shell and off-shell kaon TMD at \( p^2 = 0.6 \) GeV$^2$. In the on-shell case, \(f(x,\bm{k}_{\perp}^2)\) peaks as \(x \rightarrow 0\), differing from the pion on-shell TMD. The off-shell TMD shows a stronger $x$-dependence, similar to the pion off-shell TMD, but its peak shifts to \(x < 0.5\), in contrast to the symmetry about $x=0.5$ seen in both on-shell and off-shell pion TMDs. Integrating over \( \bm{k}_{\perp} \) recovers the off-shell kaon PDF given in Equation~(\ref{hpdf}).

Here we consider only the off-shell $u$ quark TMD, since the off-shell $s$ quark TMD can be obtained from that of the $u$ quark according to Ref.~\cite{Shi:2020pqe}.
\begin{align}\label{aG91}
 f^u(x,\bm{k}_{\perp}^2,p^2)= f^s(1-x,\bm{k}_{\perp}^2,p^2).
\end{align}

\section{Summary and Outlook}\label{excellent}
We study the off-shell generalized parton distributions (GPDs) and transverse momentum dependent parton distributions (TMDs) of the kaon in the Nambu–Jona-Lasinio (NJL) model using proper time regularization. The corresponding off-shell form factors (FFs), parton distribution functions (PDFs), and impact parameter dependent PDFs are derived and systematically compared with their on-shell counterparts.


Unlike on-shell GPDs, off-shell GPDs lack crossing symmetry, leading to Mellin moments that contain both even and odd powers of the skewness parameter. This gives rise to new off-shell FFs. Our results show clear modifications in kaon GPDs due to off-shell effects. Certain properties valid for on-shell GPDs—such as symmetry constraints and polynomiality conditions—may no longer hold in the off-shell case.


The relative off-shell effect in kaon GPDs is found to be between about $10 \%$ and $25 \%$. A comparison with pion off-shell GPDs is also carried out. From the Mellin moments of the kaon GPDs, we extract the off-shell FFs and gravitational form factors (GFFs), and further compare the kaon off-shell FFs with those of the pion.



{We also examine the off-shell TMD of the kaon, which shows a stronger dependence on $x$—a feature shared with the pion off-shell TMD. However, unlike the pion off-shell TMD, which is symmetric about \(x=0.5\), the peak position of the kaon off-shell TMD shifts to \(x<0.5\), marking a clear distinction between the two. }


In summary, the NJL model proves effective in describing the off-shell structure of pions and kaons. Future work may extend these calculations to off-shell GPDs and FFs of vector mesons (e.g., $\rho$ and $K^{*}$) and nucleons. Repeating the present analysis using models with more realistic interactions could also offer further insight into the \mbox{underlying dynamics}.


\acknowledgments
Work supported by: the Scientific Research Foundation of Nanjing Institute of Technology (Grant No. YKJ202352), the Natural Science Foundation of Jiangsu Provincial Department of Education (Grant No. 25KJD140001)

\appendix
\section{APPENDIX 1: USEFUL FORMULAE}
Here we use the gamma-functions ($n\in \mathbb{Z}$, $n\geq 0$)
\begin{subequations}\label{cfun}
\begin{align}
\mathcal{C}_0(z)&=\int_0^{\infty} ds\, s \int_{\tau_{uv}^2}^{\tau_{ir}^2} d\tau  e^{-\tau (s+z)}\nonumber\\
&=z[\Gamma (-1,z\tau_{uv}^2 )-\Gamma (-1,z\tau_{ir}^2 )]\,, \\
\mathcal{C}_n(z)&=(-)^n\frac{\sigma^n}{n!}\frac{d^n}{d\sigma^n}\mathcal{C}_0(\sigma)\,, \\
\bar{\mathcal{C}}_i(z)&=\frac{1}{z}\mathcal{C}_i(z).
\end{align}
\end{subequations}
where $\tau_{uv,ir}=1/\Lambda_{\text{UV},\text{IR}}$ are, respectively, the infrared and ultraviolet regulators described above.

The functions denoted by $\sigma$ are defined as follows:
{\allowdisplaybreaks
\begin{subequations}\label{cf}
\begin{align}
\sigma_1&=(1-x)M_u^2+xM_s^2-x(1-x)p^2\,, \\
\sigma_2&=(1-x)M_u^2+xM_s^2-x(1-x)p^{\prime 2}\,, \\
\sigma_3&=M_u^2-x(1-x)t\,, \\
\sigma_4&=\frac{1-x}{1+\xi}M_u^2+\frac{x+\xi}{1+\xi}M_s^2-\frac{x+\xi}{1+\xi}\frac{1-x}{1+\xi} p^2\,, \\
\sigma_5&=\frac{1-x}{1-\xi} M_u^2+\frac{x-\xi}{1-\xi}M_s^2-\frac{x-\xi}{1-\xi} \frac{1-x}{1-\xi} p^{\prime 2}\,, \\
\sigma_6&=M_u^2-\frac{1}{4}(1+\frac{x}{ \xi })(1-\frac{x}{\xi }) t\,, \\
\sigma_7&=(1-\alpha)M_u^2+\alpha M_s^2\nonumber\\
&-\alpha \left(\left(\frac{\xi-x}{2\xi}+\alpha\frac{1-\xi}{2\xi}\right) p^2+\left(\frac{\xi+x}{2\xi}-\alpha\frac{1+\xi}{2\xi}\right)p'^2\right)\nonumber\\
&-\left(\frac{\xi+x}{2\xi}-\alpha \frac{1+\xi}{2\xi}\right) \left(\frac{\xi-x}{2\xi}+\alpha \frac{1-\xi}{2\xi}\right) t\,, \\
\sigma_8&=x(x-1)p^{\prime 2}+y(y-1)p^2+xy(p^2+p^{\prime 2})\nonumber\\
&+(x+y)M_u^2+(1-x-y)M_s^2\,, \\
\sigma_9&=(1-x)M_u^2+xM_s^2+(1-\alpha-x) \alpha \bm{q}_{\perp}^2\nonumber\\
&-(x\alpha p^2+x(1-\alpha-x) p^{\prime 2})\,, \\
\sigma_{10}&=\bm{k}_{\bot }^2+(1-x)M_u^2+xM_s^2-x(1-x)p^2,
\end{align}
\end{subequations}

\bibliographystyle{apsrev4-1}
\bibliography{zhang}


\end{document}